\documentclass[]{jfm}

\usepackage{graphicx}
\usepackage{psfrag}
\usepackage{braket}
\usepackage{natbib}
\usepackage{color}
\definecolor{grey}{rgb}{0.5,0.5,0.5}

\ifCUPmtlplainloaded \else
  \checkfont{eurm10}
  \iffontfound
    \IfFileExists{upmath.sty}
      {\typeout{^^JFound AMS Euler Roman fonts on the system,
                   using the 'upmath' package.^^J}%
       \usepackage{upmath}}
      {\typeout{^^JFound AMS Euler Roman fonts on the system, but you
                   dont seem to have the}%
       \typeout{'upmath' package installed. JFM.cls can take advantage
                 of these fonts,^^Jif you use 'upmath' package.^^J}%
      }
  \else
  \fi
\fi

\ifCUPmtlplainloaded \else
  \checkfont{msam10}
  \iffontfound
    \IfFileExists{amssymb.sty}
      {\typeout{^^JFound AMS Symbol fonts on the system, using the
                'amssymb' package.^^J}%
       \usepackage{amssymb}%
         \let\leq=\leqslant
         
      }{}
  \fi
\fi

\ifCUPmtlplainloaded \else
  \IfFileExists{amsbsy.sty}
    {\typeout{^^JFound the 'amsbsy' package on the system, using it.^^J}%
     \usepackage{amsbsy}}
    {\providecommand\boldsymbol[1]{\mbox{\boldmath $##1$}}}
\fi

\newcommand\Rey{\mbox{\textit{Re}}}  

\newsavebox{\astrutbox}
\sbox{\astrutbox}{\rule[-5pt]{0pt}{20pt}}

\title{Subcritical versus supercritical transition to turbulence in curved pipes}

\author[J.\ K\"{u}hnen, P.\ Braunshier, M.\ Schwegel, H.\ C.\ Kuhlmann and B.\ Hof]%
{J.\ns K\ls \"{U}\ls H\ls N\ls E\ls N\ls $^1$%
  \thanks{Email address for correspondence: jakob.kuehnen@ist.ac.at},\ns
P.\ns B\ls R\ls A\ls U\ls N\ls S\ls H\ls I\ls E\ls R\ls $^2$,
M.\ns S\ls C\ls H\ls W\ls E\ls G\ls E\ls L\ls $^2$,
H.\ns C.\ns K\ls U\ls H\ls L\ls M\ls A\ls N\ls N$^2$ 
\and B.\ns H\ls O\ls F\ls $^1$}

\affiliation{
$^1$IST Austria, Am Campus 1, A-3400 Klosterneuburg\\[\affilskip]
$^2$Institute of Fluid Mechanics and Heat Transfer, Vienna University of Technology,
Getreidemarkt 9, A-1060 Vienna}

\pubyear{2014}
\volume{xxx}
\pagerange{xxx--xxx}
\date{?; revised ?; accepted ?.}

\graphicspath{{./figures/}}

\begin{document}

\maketitle

\begin{abstract}
Transition to turbulence in straight pipes occurs in spite of the linear stability of the laminar Hagen--Poiseuille flow if the amplitude of flow perturbations as well as the Reynolds number exceed a minimum threshold (subcritical transition). As the pipe curvature increases centrifugal effects become important, modifying the basic flow as well as the most unstable linear modes. If the curvature (tube-to-coiling diameter $d/D$) is sufficiently large a Hopf bifurcation (supercritical instability) is encountered before turbulence can be excited (subcritical instability). We trace the instability thresholds in the $\Rey-d/D$ parameter space in the range $0.01\leq\ d/D \leq0.1$ by means of laser-Doppler velocimetry and determine the point where the subcritical and supercritical instabilities meet. Two different experimental setups were used: a closed system where the pipe forms an axisymmetric torus and an open system employing a helical pipe. Implications for the measurement of friction factors in curved pipes are discussed.
\end{abstract}



\section{Introduction}\label{sec:intro}

Past studies of flows in curved pipes have been largely motivated by the need for reliable friction-factor data. The pressure drop in pipes is a fundamental design parameter in industry and the accurate prediction of the drag is of importance for many practical applications. For that reason friction-factor correlations for curved pipes are abundant in literature, see e.g.\ \cite{Naphon2006} and \cite{Vashisth2008} for corresponding compilations. The dependence of the friction factor on the Reynolds number and change in slope encountered with increasing $\Rey$ have commonly been used to estimate the critical threshold for the onset of turbulence. The dependence of the critical Reynolds number on pipe curvature derived in this way is thus entirely based on empirical relations found from pressure-drop measurements.

For straight pipes it is well established that the laminar Hagen--Poiseuille flow is linearly stable at all flow speeds and transition to turbulence results from finite-amplitude perturbations often referred to as sub-critical transition \citep{Grossmann2000}. Once triggered the transition to turbulence can be described as catastrophic, i.e., it is abrupt and distinct concerning flow properties such as the pressure drop. Hence, the distinction between laminar and turbulent flow in a straight pipe is clear cut, in principle. A complication, however, is caused by spatio-temporal intermittency. The co-existence of patches of laminar and turbulent flow in the transitional regime \citep{Wygnanski1973,Nishi2008,Samanta2011} and the absence of a linear instability has caused considerable difficulties in understanding the transition and in defining a threshold Reynolds number for the \textit{onset of turbulence} \citep{Eckhardt2007,Mullin2011}. While intermittent turbulent spots are transient at low Reynolds numbers, turbulence becomes sustained at larger Reynolds numbers when the spreading rate of turbulent spots outweighs their decay rate. \cite{Avila2011} have shown both rates to coincide at a critical Reynolds number $\Rey\cong2040$ which provides a natural criterion for defining the onset of turbulence.

The flow in curved pipes is further complicated by centrifugal forces. The imbalance between centrifugal forces and the cross-stream pressure gradient leads to a secondary flow in form of a pair of steady streamwise Dean vortices symmetric with respect to the equatorial plane. Associated with the Dean vortices the maximum of the streamwise velocity is shifted radially from the centre of the pipe towards the outer wall. Due to the increased cross-stream gradient of the streamwise velocity near the outer wall the drag in curved pipes is considerably higher when compared to straight pipes. This effect is more pronounced in laminar flows, while the mean wall shear stress in turbulent flows is less affected. Therefore, the difference in pressure drop between laminar and turbulent flow is much smaller than in straight pipes. \cite{Cioncolini2006}, hereinafter referred to as C06, reported a small range of Reynolds numbers in the transitional regime where the laminar pressure drop is even larger than the turbulent one.

\begin{figure}
  \centerline{\includegraphics[clip,width=0.94\textwidth,angle=0]{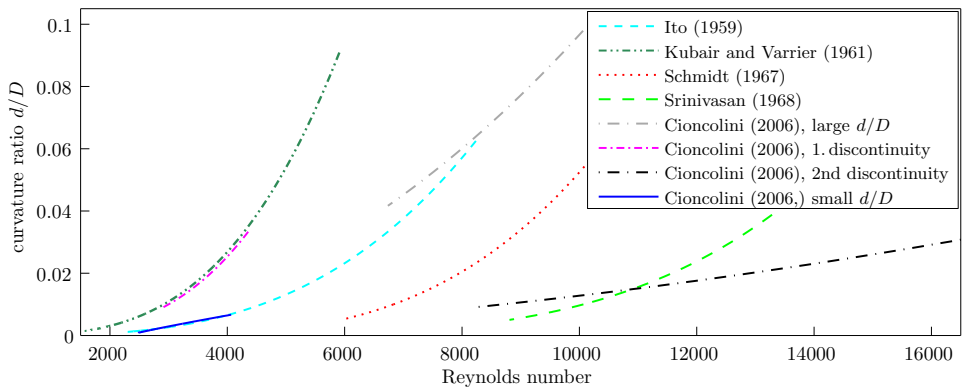}}
  \caption{\label{fig:CorrelfromLit}Critical Reynolds numbers as functions of the curvature for selected publications. The data were taken from table 4 of \cite{Vashisth2008}.}
\end{figure}

Figure \ref{fig:CorrelfromLit} shows critical Reynolds numbers for the transition to turbulence as a function of the curvature from several selected publications (adopted from \cite{Vashisth2008}, table 4) based on pressure drop measurements and observations of discontinuities in the $log(f)-log(\Rey)$-plane respectively. These critical Reynolds numbers scatter over a wide range. As a common trend, however, the critical Reynolds number increases strongly with increasing curvature.

The similar approach of all investigators using pressure drop measurements shall be exemplified considering the most recent investigation by C06. They tested twelve helically coiled pipes with curvatures $0.003\leq d/D \leq 0.14$ made of copper tubes wound around a cylinder. The inlet section consisted of a straight inlet pipe of one hundred pipe diameters followed by one revolution of the coiled pipe. The coiled part of the inlet was intended to damp entrance effects and to yield a fully developed flow. Downstream of the inlet the pressure drop over 0.25 to 1 revolutions of the coiled pipe was measured. The friction factor profiles obtained were analyzed to determine the critical Reynolds number. Based on the observations of discontinuities in the $log(f)-log(\Rey)$-plane C06 defined unique critical Reynolds numbers for small and large curvatures and two different critical Reynolds numbers for intermediate curvatures all of which are displayed in figure \ref{fig:CorrelfromLit} as four curve segments.

Recalling the above mentioned intermittent regime in straight pipes, where turbulent spots coexist with the laminar flow, the existence of such intermittent regime seems also probable at least for mildly curved pipes. However, integral pressure-drop measurements are insufficient to investigate intermittent flows. Furthermore, the observation of a supercritical instability spuriously and incoherently reported by earlier investigations of \cite{Sreenivasan1983,Webster1993,Webster1997} and \cite{Piazza2011} requires a space- and time-resolved measurement of the flow field in order to shed more light on the laminar-turbulent transition in curved pipes.

Experiments by \cite{Kuehnen2014}, hereinafter referred to as K14, have already revealed a multi-stage transition to turbulence in a toroidal pipe with $d/D=0.049$ by means of stereoscopic particle image velocimetry (S-PIV). They identified and characterized distinct flow states. The onset of time dependent flow could be determined at a first critical Reynolds number of $\Rey_{c1}=4075\pm2\%$. The bifurcation was found to be of supercritical Hopf type, leading to a wave traveling along the pipe. At $\Rey_{c2}\approx4400$ a secondary Hopf bifurcation occurs. After the secondary bifurcation the flow becomes chaotic as signified by a broad peak in the power spectrum. Fully turbulent flow with the characteristic friction factor scaling and the change in the slope of the friction factor in the $log(f)-log(\Rey)$-plane was obtained for $\Rey\gtrsim8000$. Hence, measurements of the pressure-drop fail in detecting the onset of the first instability to time dependent flow ($\Rey_{c1}$).

In view of the lack of detailed results on transitional curved pipe flow and the qualitative and quantitative differences among previous investigations a systematic variation of the curvature is desirable. From previous investigations one can suppose the Hopf-bifurcation point to move to larger Reynolds numbers as the curvature decreases with $\Rey_{c1}\to \infty$, as straight pipe flow is linearly stable. On the contrary, the subcritical transition to turbulence is assumed to arise at lower Reynolds numbers as the curvature decreases, because it can be expected to approach $\Rey\cong2040$ for straight pipes. Hence, we expect a range of curvature at which the supercritical instability and the sub-critical transition to turbulence compete. This regime is targeted in the present investigation. We focus on the linear instability threshold from laminar to oscillatory Dean flow (traveling waves) and on the onset of subcritical turbulence for small and moderate curvatures. The instabilities are detected experimentally by means of laser-Doppler velocimetry and gathered in the Reynolds number -- curvature plane. To this end we used two different experimental setups: a closed system consisting of a toroidal pipe and an open system consisting of a helical pipe with small pitch.

\section{Experimental setup and methods of investigation}\label{sec:experimenalsetup}

Five different toroidal pipes and nine different helical pipes all with curvatures in the range $0 < d/D \leq 0.1$ are used to investigate the transition to turbulence of the fully developed laminar flow. The geometrical properties of these pipes are listed in table \ref{tab:overviewdesignparam}. For reference in the text each particular pipe is assigned with a shortcut where T stands for toroidal pipe (nos.\ 1--5) and H for helical pipe (nos.\ 6--14) in conjunction with the respective curvature.

\begin{table}
\centering
\caption{\label{tab:overviewdesignparam}Overview of the 14 different experimental setups and their geometric properties. Pipes are designated using T for a toroidal pipe (no.\ 1--5) and H for a helical pipe (no.\ 6--14), combined with the curvature $d/D$. Pipe T-0.049 and respective results are from K14.}
\medskip
\begin{tabular}{@{\extracolsep{1.4ex}}rllll} 
	Nos.  & designator & $d$ [mm] & $D$ [mm] & $d/D$ \\ \hline
	1 & T-0.1 & 30.3 & 303 & $0.1$ \\
	2 & T-0.075 & 30.3 & 404 & $0.075$ \\
	3 & T-0.049 & 30.3 & 614 & $0.049$ \\
	4 & T-0.041 & 15.2 & 370.7 & $0.041$ \\
	5 & T-0.034 & 15.2 & 447 & $0.034$ \\
	6 & H-0.082 & 10.2 & 124.7 & $0.082$ \\
	7 & H-0.072 & 10.2 & 141 & $0.072$ \\
	8 & H-0.062 & 10.2 & 165.4 & $0.062$ \\
	9 & H-0.051 & 10.2 & 198.5 & $0.051$ \\
	10 & H-0.041 & 10.2 & 249 & $0.041$ \\
	11 & H-0.031 & 10.2 & 328.3 & $0.031$ \\
	12 & H-0.026 & 10.2 & 399 & $0.026$ \\
	13 & H-0.021 & 10.2 & 497 & $0.021$ \\
	14 & H-0.01 & 10.2 & 1017 & $0.01$ \\  \hline
\end{tabular}
\end{table}

Both types of setups have specific advantages and disadvantages. The closed toroidal pipe is geometrically very precise, has zero pitch and is better suited for full optical access. But the finite length of the toroidal system ($\pi D$) and the boundary conditions associated with the driving realized by an actuated rolling sphere (see section \ref{subsec:toroidalsetup}) can potentially affect the transition. Due to this boundary condition only the flow in a certain azimuthal range can be considered fully developed. For T-0.049 this range is limited to $\varphi \in [0.85\pi,1.15\pi]$, where $\varphi$ is the toroidal angle and $\varphi=0$ corresponds to the location of the actuated sphere (see K14). The boundary condition also restricts the usable curvature. If the curvature is too large the flow in the whole toroidal pipe may be affected by the driving mechanism.

The open helical pipes are realized using stiff hydraulic hoses (see section \ref{subsec:helicalsetup}). By bending into helical pipes the curvature can be varied with relative ease. Although great care was taken for setting up each curvature the system is intrinsically less accurate than the precisely machined rigid torus, since the cross section of the pipe may slightly be affected by the curvature. In addition, the helical setup has a small pitch, i.e.\ an increase in coil elevation per revolution which gives rise to an additional torsional force on the flow. However, the effect of torsion on the flow is considered negligible provided the coil pitch is lower than the coil radius (toroidal approximation for helical coils), see e.g.\ \cite{Germano1982}, \cite{Yamamoto1994} and \cite{Huttl2000}. As an advantage, the helical setup allows for a longer continuous investigation of the flow, not interrupted by the passage of the driving mechanism. Additionally, the inlet flow from a straight pipe can be kept laminar to very high Reynolds numbers and the length of the helical pipe can be made quite long to fully rule out any entrance effects. Since helical pipes represent the quasi-standard for experimental investigations of curved pipe flow, the investigation of the transition in helical as well as in toroidal geometries provides an important consistency check.

\subsection{Toroidal pipe}\label{subsec:toroidalsetup}

\begin{figure}
  \centerline{\includegraphics[clip,width=0.54\textwidth,angle=0]{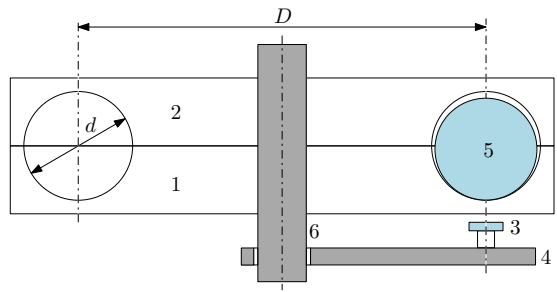}}
  \caption{\label{fig:Torus-Sketch}Sketch (side view) of the setup with toroidal pipes. Lower (1) and upper (2) plexiglass disc are bolted together and sealed using rubber O-rings (screws and sealings not shown) to ensure leakproof tightness. The two discs are machined such that they realize a toroidal pipe with internal diameter $d$ and toroidal diameter $D$. A flat pot magnet (3) with threaded stem is incorporated to the boom (4) right below the centreline of the tube to move a ferromagnetic sphere (5) placed into the tube. The boom is rotated around the shaft (6) by a geared direct current motor (not shown). Drawing not to scale. }
\end{figure}

The experimental setup employing a toroidal pipe is realized by a stationary torus (toroidal cavity) made from perspex. Figure \ref{fig:Torus-Sketch} shows a sketch of the main parts of the facility. For manufacturing reasons and to provide access to the cavity the top and bottom halves of the torus are machined into perspex plates which are accurately assembled to form the closed toroidal cavity. To drive the fluid motion a ferromagnetic stainless chromium steel sphere with diameter slightly less than the toroidal tube is placed into the toroid in addition to the working fluid. The steel sphere is actuated from outside the toroidal cavity using a strong permanent magnet mounted on a rotating boom. To achieve a constant and precisely adjustable flow rate in the torus the boom is rotated at a constant angular velocity, thereby steadily moving the sphere inside the torus driving the fluid. This allows to precisely adjust the flow rate in the torus. The driving system to rotate the shaft consists of an electric gear motor combined with a belt drive. The Reynolds number is defined as $\Rey = U d/\nu$, where $d$ is the diameter of the tube, $\nu$ the kinematic viscosity of the fluid, and the bulk velocity $U = \Omega D/2$ with angular velocity $\Omega$ of the rotating boom and $D$ the diameter of the centre circle of the torus. Further details can be found in K14.

\subsection{Helical pipe}\label{subsec:helicalsetup}

To detect a Hopf bifurcation in the range of Reynolds numbers covered by previous investigations (figure \ref{fig:CorrelfromLit}) it is essential to realise a laminar flow. Thus perturbations of the basic laminar flow must be minimised to avoid a premature onset of turbulence. For that reason the helical pipe has been set up similar to straight-pipe-flow facilities which have proven an excellent performance in this respect \citep[see e.g.][]{Hof2006,Kuik2010}.

\begin{figure}
  \centerline{\includegraphics[clip,width=0.82\textwidth,angle=0]{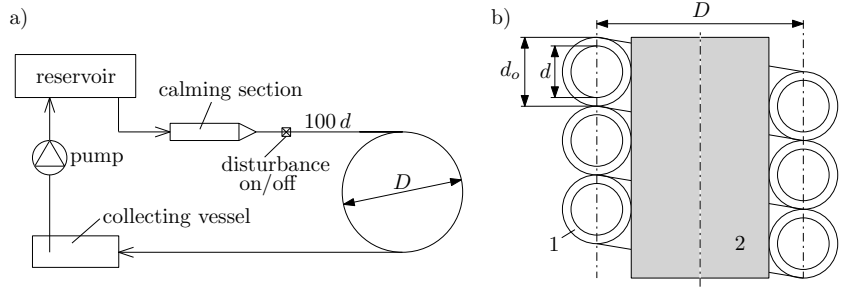}}
  \caption{\label{fig:Helix-Sketch}(a) Block diagram of the setup with helical pipes and (b) sketch (side view) of the helical pipe (1) with inner diameter $d$ and outer diameter $d_o$. The pipe is wound around a cylinder (2) of diameter $D_C$. The number of windings varied between each particular setup, because the total length of the coiled pipe ($1050\,d$) was kept constant. All measurements were taken at $1000\,d$ downstream of the inlet to the coiled section. Drawings are not to scale.}
\end{figure}

Figure \ref{fig:Helix-Sketch}a provides a schematic overview of the main parts of the helical-pipe facility: Coming from a reservoir the fluid driven by a constant pressure head first enters a calming section. The flow rate of the system can be adjusted manually by changing the total pressure head. The calming section consists of a settling chamber, honeycombs, and a series of coarse and fine screens to reduce any undesired disturbances in the flow before entering the pipe through a smoothly contracting nozzle. After a straight entrance length of $100\,d$, where the flow is proven to develop a laminar profile up to $\Rey=8000$, the flow enters a coiled section with coiling diameter $D$. The transition from the straight to the coiled section is quasi-abrupt but smoothed due to the stiff and continuous hose used. In the coiled section the hose is wound several times around a cylinder of appropriate diameter $D_C$ for a total length of $1050\,d$. For experiments in which a turbulent inflow to the helically coiled section is desired a continuous disturbance can be activated in the straight pipe right after the calming section. To determine the Reynolds number the flow rate is measured manually by weighing the mass per time at the exit into a collecting vessel. Furthermore, the temperature of the water was constantly measured to determine the viscosity of the fluid. The fluid is then pumped back into the header tank. All experiments were performed with water as the test fluid at ambient temperature.

Figure \ref{fig:Helix-Sketch}b shows a cross section of the coiled section. It consists of a long hydraulic hose (TU1610C by SMC) with $d=10.2\pm 0.15$\,mm internal diameter and $d_o=16\pm 0.1$\,mm outer diameter. The hose is made of polyurethane, providing a transparent, rather stiff yet sufficiently flexible tube. The diameter of the cylinder is $D_C$, the diameter $D$ of the coiling (equal to the diameter of the centre circle of the torus) is $D = D_C + d_o$. $D_C$ is varied from 108.7\,mm for the largest curvature to 1001\,mm for the smallest curvature. The accuracy of the Reynolds number $\Rey$ was determined to be $\pm 3$\%.

\subsection{Measurements}\label{subsec:measurements}

A 2D-LDV system (DANTEC Dynamics) was used to record velocity time-series of the streamwise (azimuthal) velocity component $w$. Subsequently $w$ was Fourier analyzed for different Reynolds numbers. The dimensionless frequency (Strouhal number) is obtained as $\hat{f} = fd/U$, where $f$ is the frequency in Hz. A cartesian ($x,y,z$) coordinate system is used in the meridional observation plane at constant toroidal (helical) angle $\varphi$. The origin of coordinates is set in the centre of the pipe, the $x$-axis being directed radially towards the inside of the coiling and the $z$-axis pointing in the streamwise direction. The respective Cartesian velocity components are denoted $(u,v,w)$.

As shown by \cite{Webster1997} and K14 the first instability is an oscillatory mode with the largest amplitude at two points located mirror-symmetric with respect to the equatorial plane $y=0$ in the inner half of the curved pipe. K14 have shown that the spatial distribution of the fundamental Fourier mode $A_1(x,y)$ with frequency $f_1$ and the corresponding streamwise velocity fluctuation has a maximum at $(x,y)\cong(0.17\,d,\pm0.31\,d)$ for a curvature of $d/D=0.049$. Since the best signal-to-noise ratio for detecting the oscillatory perturbations is achieved at these locations, the flow velocity was measured at $(x,y)=(0.17\,d,0.31\,d)$. However, the optimum locations for other curvatures may differ slightly. In the helical setup the measurement plane was always located $1000\,d$ downstream of the inlet to the coiled section to rule out any transient effects.

For $d/D=0.049$ K14 have found the first instability at $\Rey_{c1}$ to be oscillatory with a single dominant frequency $\hat{f_1}$. Upon an increase of the Reynolds number the amplitude of the fundamental mode as well as weak higher harmonics increase continuously. A second oscillatory instability was detected at $\Rey_{c2}$, signified by an ambiguous peak-broadening of $\hat{f_1}$ which could not be safely distinguished from an additional fundamental frequency slightly different to $\hat{f_1}$ in the power spectrum. Further beyond the second critical point additional frequencies in the vicinity of $\hat{f_1}$ have been detected. The same fundamental frequency $\hat{f_1}$ is still dominant, but its amplitude decreases. Since we expect a similar scenario also for other curvatures, we use these indications to detect $\Rey_{c1}$ and $\Rey_{c2}$. I.e., the onset of a single dominant frequency $\hat{f_1}$ is indicative of $\Rey_{c1}$ and the onset of peak-broadening of $\hat{f_1}$ and the emergence of additional frequencies is indicative of $\Rey_{c2}$.

\begin{figure}
  \centerline{\includegraphics[clip,width=0.94\textwidth,angle=0]{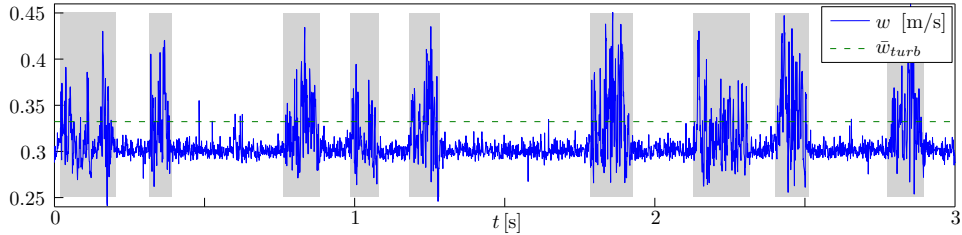}}
  \caption{\label{fig:lam-turb-intermittency}Streamwise velocity $w$ measured in H-0.02 at $(x,y)=(0.17\,d, 0.31\,d)$ for $\Rey=4140$. The signal shows an intermittent behavior indicative of the subcritical scenario. Turbulent regions are indicated by a gray background.}
\end{figure}

If instead of a continuous supercritical bifurcation a subcritical, direct transition to turbulence occurs the LDV signal is clearly different. Although turbulent flow at the inner side of curved pipes (i.e.\ the measurement location) was reported to be damped considerably compared to the outer side \citep{Sreenivasan1983,Noorani2013}, in our measurements the onset of turbulence could be distinctly identified either by an increase of the standard deviation by a factor of $\sim4$ or by an increase of the average velocity as indicated by $\bar{w}_{turb}$ in figure \ref{fig:lam-turb-intermittency}. The figure shows the intermittent regime in a mildly curved pipe (H-0.02) qualitatively similar to observations in straight pipes, where laminar and turbulent flows can coexist. Since the time scales at the onset of turbulence can be extremely large \citep{Avila2011} we select the Reynolds number $\Rey_{t}$ where the turbulence fraction reaches 50\% (measured $1000\,d$ after the entrance to the coil) as the criterion for the onset of turbulence. Note that this threshold will be somewhat above the onset of sustained turbulence. It is however much more convenient to determine and provides a sufficiently accurate estimate for the purpose of the present investigation.

\section{Results}\label{sec:results}

\begin{figure}
  \centerline{\includegraphics[clip,width=0.98\textwidth,angle=0]{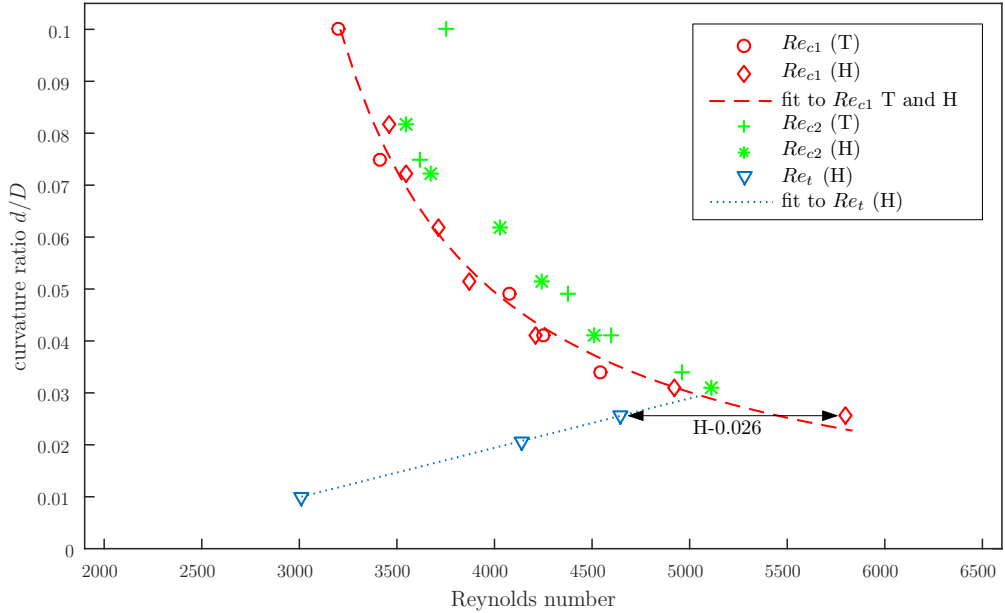}}
  \caption{\label{fig:Curvature-vs-Re}Measured bifurcation points $\Rey_{c1}$ and $\Rey_{c2}$ and Reynolds number $\Rey_{t}$ where the turbulence fraction reaches 50\% in toroidal (T) and helical (H) pipes.}
\end{figure}

For the five toroidal pipes investigated we find a supercritical Hopf bifurcation from the steady basic flow to an oscillatory flow at $\Rey_{c1}$ as the Reynolds number is increased quasi-statically (circles in figure \ref{fig:Curvature-vs-Re}). Further increasing the Reynolds number beyond $\Rey_{c1}$ the oscillation amplitude grows continuously from zero at the threshold. At some point the amplitude of the single dominant frequency reaches a maximum as function of $\Rey$ and begins to decrease on an increase of $\Rey$. At this second critical Reynolds number  $\Rey_{c2}$ (plusses in figure \ref{fig:Curvature-vs-Re}) also the blurred peak-broadening sets in and additional frequencies arise in the spectrum very close to the fundamental frequency as observed by K14.

For the six helical pipes with curvatures $0.031\leq d/D \leq0.082$ we similarly find a supercritical Hopf bifurcation at $\Rey_{c1}$ (diamonds in figure \ref{fig:Curvature-vs-Re}) followed by $\Rey_{c2}$ (stars) at a further increase. Note that in this curvature regime $\Rey_{c1}$ is found independently of whether the inlet flow originating from the straight pipe is laminar or turbulent.

As can be seen from figure \ref{fig:Curvature-vs-Re} $\Rey_{c1}$ monotonically increases as $d/D\to 0$ and the results for $\Rey_{c1}$ from the helical pipes are in excellent agreement with the measurements employing the toroidal pipes. For e.g.\ $d/D=0.041$, the only curvature equally investigated in both setups (T-0.041 and H-0.041), the difference is $\Delta \Rey_{c1}=87$. This difference is well within the accuracy by which the Reynolds number could be determined (see section \ref{subsec:toroidalsetup} and \ref{subsec:helicalsetup}). Except for the relatively large curvature $d/D=0.1$ the second critical Reynolds number $\Rey_{c2}$ is also increasing for $d/D\to 0$, although no clear trend could be detected for the dependence of the distance between the two critical points $\Delta \Rey = \Rey_{c2} - \Rey_{c1}$ on $d/D$. However, also the results for $\Delta \Rey$ seem to be consistent between the toroidal and helical pipes.

For the three helical pipes with the smallest curvatures (H-0.026, H-0.021 and H-0.01) on the other hand we could not detect a supercritical bifurcation. Instead, subcritical transitions are found at $\Rey_{t}=4644$ (H-0.026), $\Rey_{t}=4141$ (H-0.021) and $\Rey_{t}=3011$ (H-0.01). The values for $\Rey_t$ are given by solid triangles in figure \ref{fig:Curvature-vs-Re}. Note that in these cases the inlet-flow into the helical pipe was turbulent. However, by reducing the finite amplitude perturbations, i.e.\ by keeping the inlet-flow into the helical pipe laminar, we were able to detect a supercritical instability at $\Rey_{c1}=5800$ for H-0.026. The two thresholds ($\Rey_{c1}$ and $\Rey_t$) for the different instabilities for this curvature are indicated in figure \ref{fig:Curvature-vs-Re} by an arrow. As the emergence of the oscillatory flow at $\Rey_{c1}=5800$ triggers turbulence $\Rey_{c2}$ could not be detected for H-0.026.

For H-0.021 and H-0.01 we could not verify a supercritical bifurcation by keeping the inlet-flow laminar, as the flow in the helical pipe could not be kept laminar sufficiently high beyond $\Rey_t$. Ink injected into the straight pipe upstream of the helical pipe for flow visualization indicated that the laminar flow coming from the straight pipe is perturbed and becomes turbulent right at the inlet section to the helical pipe, where the straight pipe quasi-abruptly turns into curved.

Assuming that $\Rey_{c1}\rightarrow\infty$ for $d/D\rightarrow0$ we can fit the data points of supercritical transition in the toroidal and helical setups to

\begin{equation}\label{fit}
    \Rey_{c1} = 77.2\, D/d + 2438
\end{equation}

Although the small number of 3 data points for $\Rey_{t}$ may seem not very reliable for the linear fit $d/D=9\cdot 10^{-6}\Rey_t-0,0185$, the point of the intersection where subcritical turns into supercritical transition depending on the curvature can be reasonably reliable conjectured at $d/D\cong0.028$.

\section{Discussion}\label{sec:discussion}

The experiments presented in this paper establish the existence of an instability threshold to an oscillatory mode (supercritical Hopf bifurcation) in the $\Rey-d/D$ parameter plane if the curvature is sufficiently large ($0.028 \leq d/D \leq 0.1$) which is masked by subcritical transition to turbulence if the curvature is small ($d/D \leq 0.028$). By reducing perturbation levels we could track and detect the supercritical bifurcation down to $d/D=0.026$.

Excellent agreement and the same Hopf bifurcation is found in the closed toroidal pipes as in the helical pipes. Hence, neither the difference in the streamwise boundary conditions nor the small pitch or the lower geometrical precision of the helical pipe has a sizable influence on the bifurcation point. The results obtained enable a better understanding of previous seemingly contradictory observations reported in the literature.

None of the previous studies of transition in curved pipes using pressure drop measurements have revealed the existence of the supercritical Hopf bifurcation and the oscillatory flow regime between $\Rey_{c1}$ and $\Rey_{c2}$. In particular, none of the correlations suggested previously even roughly captures the trend of $\Rey_{c1}$ or $\Rey_{c2}$ (for comparison see also figure \ref{fig:CorrelfromLit}). It can be concluded that not only the transition to the oscillatory part of the flow between $\Rey_{c1}$ and $\Rey_{c2}$ but also to the non-fully turbulent regime above $\Rey_{c2}$ has a negligible effect on the pressure drop.

Our measurements provide a solid basis for the existence of laminar oscillatory flow in curved pipes with critical data of unprecedent accuracy. Laminar stable traveling waves in curved pipes have only been spuriously and incoherently reported before. While \cite{Sreenivasan1983} provided more qualitative data, \cite{Webster1993,Webster1997} measured the onset of periodic low-frequency perturbation waves at $\Rey = 5060$ for $d/D=0.055$, a value which is considerably above our measurement of $\Rey_{c1}$ and even above $\Rey_{c2}$. We cannot offer a definitive explanation for this deviation. But since the first dominant frequency of the oscillatory flow is still distinctly present in the spectrum even beyond $\Rey_{c2}$ (see K14), their measurements may have been carried out in the quasiperiodic regime.

\cite{Piazza2011} numerically investigated the transition to turbulence in a torus limited to the two values of curvatures $d/D=0.1$ and $0.3$. They found a supercritical Hopf bifurcation for $d/D=0.3$. Apart from being out of the range of the present investigation the flow states found (stationary, periodic, quasi-periodic and chaotic) seem to be in qualitative agreement with our experimental results. But the supercritical Hopf bifurcation was found at $\Rey_{c1}=4575$, which is considerably above the range predicted by our fit \ref{fit}. For the curvature which is directly comparable to the present investigation ($d/D=0.1$) they found a direct transition from a steady to a quasi-periodic flow  between $5139 < \Rey < 5208$ associated with hysteresis (subcritical Hopf bifurcation). This is in contradiction to our measurements. Being aware of their results we paid high attention on possible hysteresis or a direct transition from a steady to a quasi-periodic flow in our measurements, especially for $d/D=0.1$, but did not find any indication for it. We cannot explain this discrepancy, but based on the present solid experimental evidence from two different setups we suggest that more refined simulations are needed.

As the curvature is further decreased towards the straight-pipe limit, i.e.\ for $0\leq\ d/D \leq0.028$, we observe a subcritical transition similar to straight pipes. As can be seen in comparison to figure \ref{fig:CorrelfromLit} our data points for $\Rey_{t}$ comply with the general trend of published correlations, being relatively close to \cite{Ito1959}, \cite{Kubair1961} and C06. This strongly suggests that previous investigations employing pressure drop measurements have detected the subcritical transition. Depending on the perturbation level caused by a specific setup and the respective intermittent turbulence fraction they found different scaling laws for the critical Reynolds number.

In the regime of subcritical transition the onset of turbulence is postponed compared to straight pipes and occurs at Reynolds numbers considerably larger than $\Rey=2040$ (the value where turbulence first becomes sustained in straight pipes). Our results confirm that fully developed turbulent flow emerging from a straight pipe can be completely laminarised depending on the curvature as already observed by \cite{Sreenivasan1983}. This effect is most pronounced for $d/D \cong 0.028$ where $\Rey_{c1}$(d/D) and $\Rey_t(d/D)$ intersect with laminarisation up to $\Rey \cong 5180$.

Further investigations are required to clarify the range of Reynolds numbers for which the flow can be kept laminar for $d/D<0.026$. A continuously increasing curvature at the passage from the straight to the curved inlet-section and helical pipes with higher geometric precision may enable such investigation. This would allow to track $\Rey_{c1}$ to even higher Reynolds numbers and smaller $d/D$ respectively. Finally it should be noted that existing friction factor data in the transitional regime may be somewhat inaccurate, as the effect of intermittency seems not to have been taken into account in respective pressure drop measurements.

In summary we have determined the transition threshold to unsteady flow in curved pipes for curvatures $0.01\leq\ d/D \leq0.1$ by means of laser-Doppler velocimetry. Experiments were carried out in toroidal and helical pipes and both data sets were in excellent agreement. For small curvature values a subcritical transition akin to that in straight pipes is encountered. For large curvatures on the other hand a supercritical, multi-stage transition is detected. Systematically varying the pipe curvature we have tracked the Hopf bifurcation and determined the point where the transition scenario changes from super- to subcritical ($d/D\cong 0.028$).

\vspace{0.2cm}
\paragraph{\textbf{Acknowledgements}} J.K.\ acknowledges the provision of preliminary numerical data by Philipp Schlatter and Azad Noorani of KTH Mechanics related to this work and friendly support by Gregoire Lemoult.

The project was partially funded by the European Research Council under the European Union's Seventh Framework Programme (FP/2007-2013)/ERC Grant Agreement 306589.

\bibliographystyle{jfm}

\bibliography{Literatur_JK}

\end{document}